\documentclass[12pt]{spie}  

\usepackage[colorlinks=true, allcolors=blue]{hyperref}
 
\usepackage{amsmath,amsfonts,amssymb}
\usepackage{graphicx}
\usepackage{soul}

\title{Astro 2020 State of the Profession: Astrophotonics White Paper}

\author[1*]{Pradip Gatkine}
\author[1,2,3]{Sylvain Veilleux}
\author[4]{John Mather}
\author[5]{Christopher Betters}
\author[5]{Jonathan Bland-Hawthorn}
\author[5]{Julia Bryant}
\author[4,2]{S. Bradley Cenko}
\author[6]{Mario Dagenais}
\author[1]{Drake Deming}
\author[7]{Simon Ellis}
\author[4]{Matthew Greenhouse}
\author[1]{Andrew Harris}
\author[8]{Nemanja Jovanovic}
\author[9]{Steve Kuhlmann}
\author[4]{Alexander Kutyrev}
\author[5]{Sergio Leon-Saval}
\author[10]{Kalaga Madhav}
\author[4]{Samuel Moseley}
\author[5]{Barnaby Norris}
\author[4]{Bernard Rauscher}
\author[10]{Martin Roth}
\author[1]{Stuart Vogel}

\affil[1]{Dept. of Astronomy, University of Maryland, College Park, MD, USA}
\affil[2]{Joint Space-Science Institute, Univ. of Maryland, College Park, MD, USA}
\affil[3]{Institute of Astronomy and Kavli Institute for Cosmology, University of Cambridge, Cambridge CB3 0HA, United Kingdom}
\affil[4]{NASA Goddard Space Flight Center, Greenbelt, MD, USA}
\affil[5]{Sydney Institute for Astronomy and Sydney Astrophotonic Instrumentation Labs, School of Physics, The University of Sydney, New South Wales 2006, Australia}
\affil[6]{Dept. of Electrical and Computer Engineering, Univ. of Maryland, College Park, MD, USA}
\affil[7]{Dept. of Physics and Astronomy, Macquarie University, Sydney, NSW 2109, Australia}
\affil[8]{California Institute of Technology, Pasadena, CA, USA}
\affil[9]{Argonne National Laboratory, 9700 S. Cass Avenue, Argonne, IL, USA}
\affil[10]{innoFSPEC, Leibniz-Institut f\"{u}r Astrophysik Potsdam, Germany}

\authorinfo{*Contact information: Pradip Gatkine, E-mail: pgatkine@astro.umd.edu}

\begin{document} 
\maketitle

\noindent \textbf{Executive Summary:} Astrophotonics is the application of versatile photonic technologies to channel, manipulate, and disperse guided light from one or more telescopes to achieve scientific objectives in astronomy in an efficient and cost-effective way. The developments and demands from the telecommunication industry have driven a major boost in photonic technology and vice-versa in the last $\sim$40 years.  The photonic platform of guided light in fibers and waveguides has opened the doors to next-generation instrumentation for both ground- and space-based telescopes in optical and near/mid-IR bands, particularly for the upcoming extremely large telescopes (ELTs). The large telescopes are pushing the limits of adaptive optics to reach close to a near-diffraction-limited performance. The photonic devices are ideally suited for capturing this AO-corrected light and enabling new and exciting science such as characterizing exoplanet atmospheres. The purpose of this white paper is to summarize the current landscape of astrophotonic devices and their scientific impact, highlight the key issues, and outline specific technological and organizational approaches to address these issues in the coming decade and thereby enable new discoveries as we embark on the era of extremely large telescopes. \\

\vspace{1ex}



\pagebreak

\clearpage
\setcounter{page}{1}
\pagestyle{plain}

\section{Key Issue and Overview of Impact on the Field}
\label{sec:intro}  

\subsection{Introduction}
The field of astrophotonics spans a wide range of technologies including: collecting astronomical light into guided channels (fibers/waveguides), manipulating the transport and reconfiguration of the light, and filtering/dispersing/combining the guided light. A combination of one or more of these functionalities has led to a wide spectrum of astrophotonic instruments. Just as radio astronomy finds its roots in radio communication, astrophotonics finds its roots in photonic / fiber-optic communication industry. The ongoing growth of photonics industry and astrophotonics displays a strong parallel with the development of radio communication and radio astronomy, where each positively influenced the other.  

\begin{figure} [ht]
  \begin{center}
   \begin{tabular}{c} 
   \includegraphics[height=10cm]{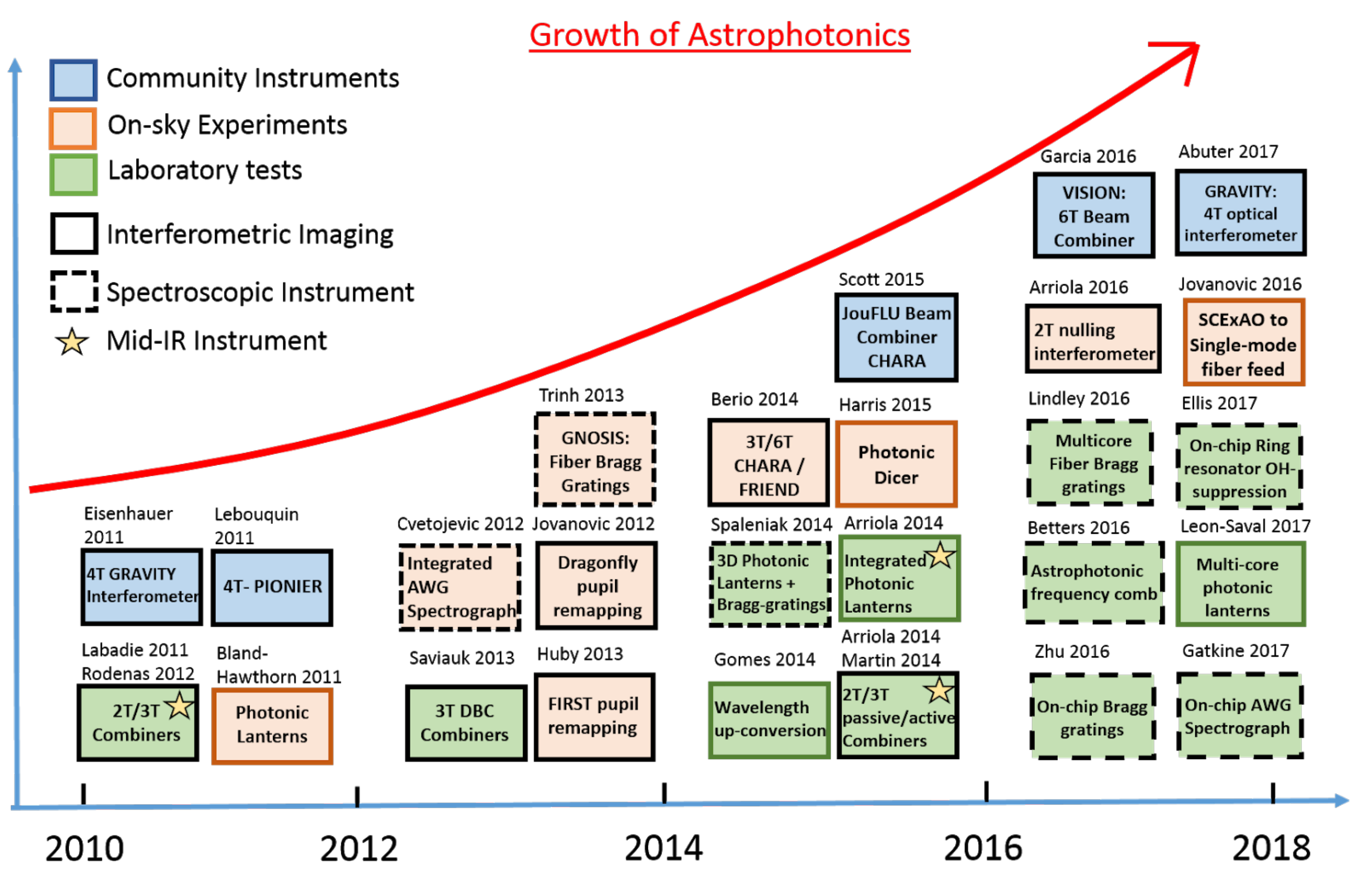}
   \end{tabular}
   \end{center}
   \caption[Picture of the spectrograph] 
   { \label{fig:Astrophotonic_growth} 
 Growth of the field of astrophotonics in the last decade}
 \end{figure}

Figure \ref{fig:Astrophotonic_growth} outlines the family of astrophotonic instruments and its superlinear growth in the last decade. The next two decades will mark the age of ELTs. Photonic technologies provide a promising platform for building next-generation instruments for the ELTs that are \textbf{flexible} (in terms of light manipulation), \textbf{compact} (volumes of a few tens of cm$^{3}$ ensuring thermal-mechanical stability) and \textbf{lightweight} (a few hundred grams), thanks to manipulation of guided light \cite{bland2009astrophotonics, allington2010astrophotonic}. In addition, they are \textbf{cost-effective}, due to their modularity and advantages of easy replication. With these inherent benefits, an accelerated development of astrophotonic instruments should be considered a key priority in the next decade. 

\noindent \ul{The key astrophotonic instruments of the last decade include:}

\noindent \textbf{1. Photonic Lanterns:} Due to the large spot size in a ground-based seeing-limited observation, channeling this light requires the use of large multimode fibers (MMFs, $\sim$several tens of modes). However, the photonic manipulation is only possible for single-mode fibers/waveguides. A photonic lantern is an adiabatic taper enabling a low-loss transition from the MMF to a set of single-mode fibers/waveguides \cite{leon2005multimode}. Photonic lanterns also make it possible to reformat the beam by rearranging the output single-mode fibers, thus tremendously enhancing the flexibility.  

\noindent \textbf{2. Complex Bragg-gratings:} The waveguides/fibers can be used as an ensemble of sharp notch filters (width  $\sim$ 1 \AA, rejection ratio $\sim$ 1000) by carefully introducing subtle refractive index variations along their length in a complex pattern \cite{bland2011complex, zhu2016arbitrary}. These devices have been demonstrated for atmospheric OH-emission suppression in the near-IR \cite{trinh2013gnosis} and are currently under development for filtering exoplanet biosignatures \cite{xie2018add}. Photonic ring resonators are also being adapted, along the same lines, to accomplish OH-emission suppression \cite{ellis2017photonic}.   

\noindent \textbf{3. Pupil remappers:} These have an on-chip 3D structure of waveguides which coherently rearranges the sub-pupils from a telescope(s) into a linear array suitable for beam combination/interferometry \cite{jovanovic2012starlight}. 

\noindent \textbf{4. Beam combiners/interferometers:} The channeling of light into waveguides allows on-chip beam combination, interferometry, and nulling of light coming from sub-pupils of one or more telescopes. The GRAVITY instrument \cite{abuter2017first} on the VLT Interferometer (VLTI) is an example of such on-chip beam combination of 4 x 8-meter telescopes in the K-band (1.95 $-$ 2.45 $\mu$m). Thanks to the ultra-fast laser inscription (ULI) methods, it has recently become possible to reliably write 3-dimensional waveguide paths in a glass substrate allowing large number ($> 10$) of baselines \cite{pedretti2018six}.

\noindent \textbf{5. Photonic spectrographs:} Various photonic implementations of dispersion have been demonstrated in the past \cite{allington2010astrophotonic, gatkine2019astrophotonic}. The most prominent technique is using an arrayed waveguide grating (AWG), which uses an on-chip phased array-like structure of waveguides to introduce progressive path lengths similar to a grating. The on-chip implementations are extremely compact and easily stackable to create an IFU or multi-object spectrograph \cite{cvetojevic2012first, harris2012applications, gatkine2017arrayed, stoll2017high, gatkine2018towards}. Such compact ($\lesssim$10 cm$^2$) photonic chips can achieve a resolution $\sim$ 20,000 in the first order.  

\noindent \textbf{6. Laser frequency combs:} The capability of optical frequency combs (Nobel Prize to Hall $\&$ H\"{a}nsch 2005) to achieve ultra-high precision of wavelength calibration on the order of 1 m/s was demonstrated with HARPS on the basis of a mode-locked laser frequency comb, connected to an atomic clock, and filtered by a cascade of Fabry-Perot etalons\cite{wilken2010high}. The goal is to improve this precision to 10 cm/s for exoplanet research. New techniques, such as four-wave-mixing have been demonstrated on-sky \cite{boggio2018wavelength}. Further photonic solutions such as micro resonators\cite{demirtzioglou2018frequency} and electro-optical modulators\cite{metcalf2019stellar} have shown the promise to lead to more affordable and stable devices, and also become useful for medium resolution spectroscopy.

Figure \ref{fig:Photonic_interferometer} shows the simplified design of a fully integrated nulling spectro-interferometer to illustrate how various astrophotonic devices come together to create an ultracompact instrument combining a slew of functionalities to enable new science. While most of the discussion in this paper is focused on ground-based telescopes, the benefits of astrophotonics are applicable to filtering, dispersion, and interferometry in space-based telescopes as well (eg: as  demonstrated in the concept Cubesat FIRST\cite{lacour2014cubesats}). 

\subsection{Science enabled by astrophotonics}

The application of photonics in telescope instrumentation has opened the doors  for new kinds of observations and this is set to expand further as the integration of photonics brings more flexibility with light manipulation. The GRAVITY instrument (K-band: 1.95-2.45 $\mu$m) is a glowing example of the synergy of bulk optics and astrophotonic instrumentation \cite{abuter2017first}. GRAVITY combines the light from 4 VLT telescopes in an integrated chip and the combined beam is then dispersed using bulk optics. With its unique and groundbreaking instrumentation, GRAVITY has led to many ‘first-of-its-kind’ observations such as: \textbf{(1)} direct detection of exoplanets as close as $\sim$100 mas \cite{lacour2019first}, \textbf{(2)} first resolution of microlensed images \cite{dong2019first}, \textbf{(3)} spatially resolved rotation of a quasar broad line region at sub-pc level \cite{sturm2018spatially}, \textbf{(4)} observation of a microquasar at sub-milliarcsecond scale resolving accretion-ejection morphology at sub-AU scale \cite{petrucci2017accretion}, \textbf{(5)} detection of orbital motions at 30-$\mu$arcsec precision near the innermost stable circular orbit of Sgr A* \cite{abuter2018a}, \textbf{(6)} detection of gravitational redshift from a star orbiting the Galactic Center massive black hole \cite{abuter2018b}, \textbf{(7)} geometric measurement of the distance to the Galactic Center with 0.3\% uncertainty \cite{abuter2019geometric}.

 \begin{figure} [ht]
  \begin{center}
   \begin{tabular}{c} 
   \includegraphics[height=11cm]{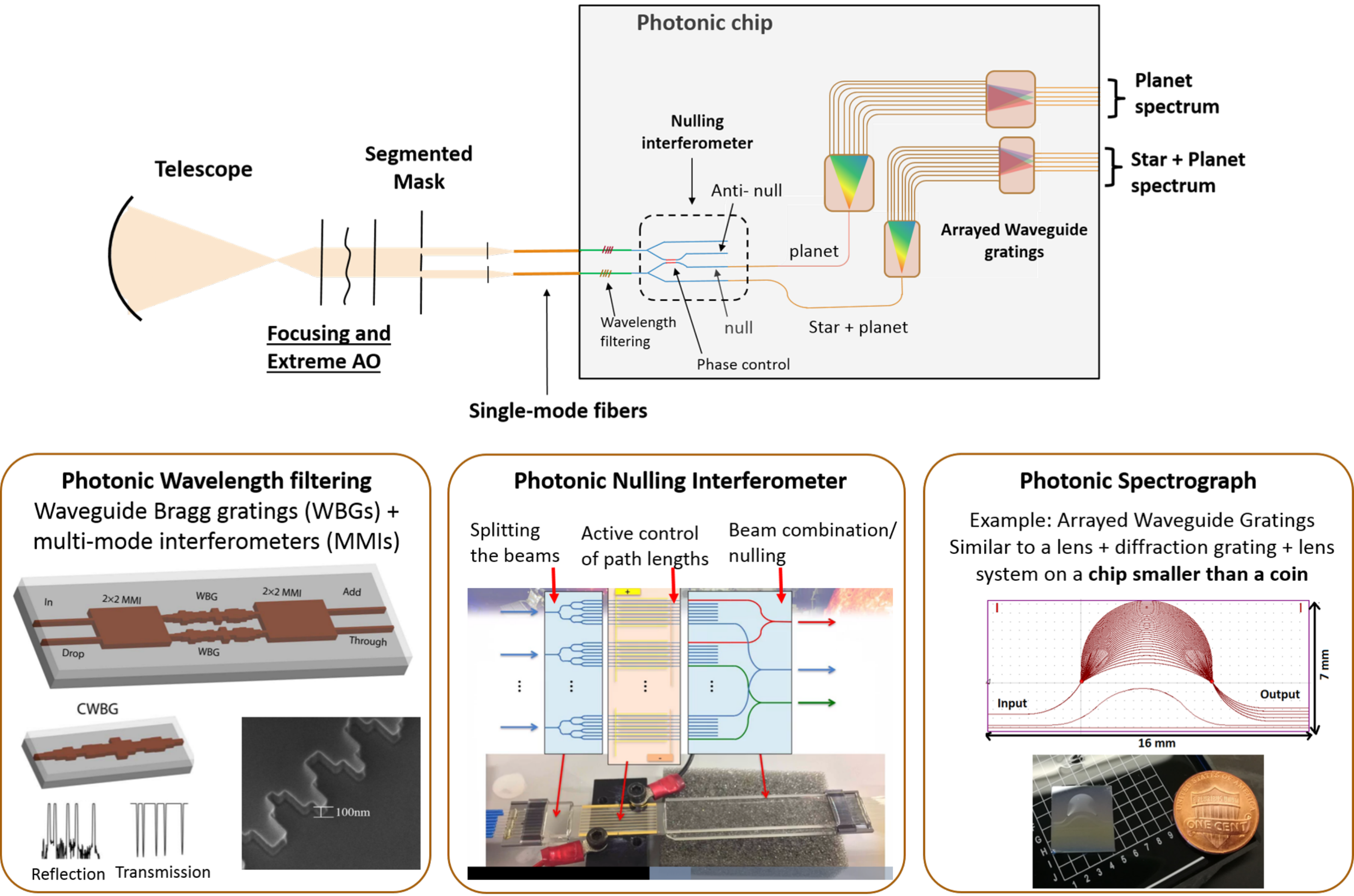}
   \end{tabular}
   \end{center}
   \caption[] 
   { \label{fig:Photonic_interferometer} 
Schematic of a fully integrated nulling spectro-interferometer on a photonic chip (top). Key techniques that have already been demonstrated on a chip (bottom row): (left) complex wavelength filtering \cite{xie2018add}, (center) pupil remapper $+$ interferometer \cite{cvetojevic2018first}, (right) photonic AWG \cite{gatkine2017arrayed}. }
 \end{figure} 

Using a similar hybrid (conventional + photonic) instrumentation, the Michigan Infrared Beam Combiner demonstrated first direct imaging of `starspots' on a star 55 pc away \cite{roettenbacher2016no}. Recent demonstration of coupling the light from Subaru extreme AO system into a single mode fiber has further strengthened the case for diffraction-limited single-mode photonic instrumentation.  With the ongoing development of astrophotonic devices, more such exciting observations are in the waiting, for example: 

\noindent \textbf{1. Direct imaging of exo-earths:} The mid-IR wavelength of $\sim$ 4-10 $\mu$m offers a `sweet spot’ where the star-planet contrast requirement is minimal for an Earth-Sun analog. Recently, chalcogenide glass-based platforms have enabled the development of a mid-IR astrophotonic nulling interferometer \cite{gretzinger2019towards}, which will bring the direct detection and characterization of exo-earths within the reach of ELTs. Further, this waveband offers signatures of important chemicals such as water, methane, and ozone. 

\noindent \textbf{2. Nulling interferometry:} Hybrid coronagraphs (fiber-fed $+$ bulk optics) have been demonstrated to achieve contrast ratio of the order of 10$^{4}$:1 \cite{haffert2018single}. With the precise beam splitting, stability, and precise phase tuning allowed by the fully-integrated, on-chip photonic nullers (see Fig \ref{fig:Photonic_interferometer}), a contrast $>$ 10$^{5}$:1 is within sight\cite{goldsmith2017improving}. Such efficient suppression of the central source (factor of $\sim$10$^{5-6}$) will enable direct imaging/spectroscopy of exoplanets and QSO hosts galaxies as shown in the schematic in Fig \ref{fig:Photonic_interferometer}. These chips are also ideally suited for high-contrast coronagraphic multi-object spectrographs for future space telescopes such as HabEx / LUVOIR \cite{coker2019}.  

\noindent \textbf{3. Precision radial velocity measurement:} A major limiting factor for achieving extremely high precision in RV is the stability of the PSF due to multimode nature of the fiber/slit feeding a conventional spectrograph. With a single-mode diffraction-limited slit in the photonic instrumentation, this problem is eliminated. In addition, a fiber etalon or an on-chip ring resonator are suitable for producing a stable frequency comb with m/s accuracy \cite{fischer2016state, bland2017mapping, ellis2017photonic}. With a multi-input AWG, it is possible to feed this calibration signal to the AWG without corrupting the astronomical source, thus allowing for an instantaneous high-precision wavelength calibration \cite{jovanovic2016enhancing, gatkine2018towards}. Such precision is crucial for detecting an exo-earth in the habitable zone around M-dwarfs. 

\noindent \textbf{4. Making large 3D maps of the early universe:} With the photonic advantage, it is possible to produce low-cost massively replicated IFU spectrographs on ELTs covering several sq. arcminutes by stacking photonic spectrographs (similar to the massively replicated VIRUS IFU concept on Hobby-Eberly Telescope \cite{hill2018virus}). In addition, the atmospheric OH-emission suppression using photonic Bragg gratings or ring resonators will reduce the background many folds (see Fig. \ref{fig:On_sky_photonics}) and achieve the faint flux limits required to study the first galaxies in the universe. 

\subsection{Impact beyond astronomy}
As described earlier, the field of astrophotonics has emerged from photonic technology that was developed for the telecommunication industry. With rising maturity and innovation in astrophotonic devices, they are now finding applications in a diverse set of fields, such as remote sensing, biophotonics, biomedical analysis, microfluidics, and the telecommunication industry itself. A few notable examples include: \textbf{a)} use of photonic lanterns in space-division multiplexing to increase the data bandwidth with each mode acting as an independent channel \cite{leon2017photonic}, \textbf{b)} application of fiber-fed IFUs and low-loss silicon-nitride (SiN) AWGs for Raman imaging $+$ spectroscopy for use in cancer diagnostics \cite{haynes2010fibre, schmalzlin2018nonscanning}, \textbf{c)} application of SiN complex Bragg gratings in medical diagnostics and other sensing applications \cite{hainberger2019silicon}, \textbf{d)} use of photonic lanterns to  enhance free-space coupling of the lasers used in LIDAR and laser communication\cite{ozdur2015photonic}.

\noindent \ul{Therefore, the resources dedicated to astrophotonics not only impact astronomy,  but they also benefit a much wider scientific and industrial community.}

\section{Strategic Plan}

\subsection{Key issues to address in the next decade}

While there has been significant in-lab and on-sky development work in astrophotonics over the last decade, there are crucial issues that need to be addressed with dedicated resources to make astrophotonic instruments science-ready and utilize their full potential. 

\noindent \textbf{1. Coupling efficiency:}  The major bottlenecks in the throughput of an astrophotonic system arises due to the coupling efficiency at the interfaces. The most lossy interfaces are: free space to (single/multi-mode) fiber, fiber to the photonic chip, and photonic chip to free space. Typical peak efficiency at each of these interfaces is $\sim$70\% which is mainly due to the mismatch of the numerical aperture (or in other words, mode profile). 

\noindent \textbf{2. Throughput:} The second most important loss is the propagation (and bending) loss in waveguides in a photonic chip. This loss can be minimized by optimizing the design and the index contrast between the waveguide core and cladding.  

\noindent \textbf{3. Polarization:} This problem only exists in on-chip devices. The on-chip waveguides are typically rectangular or there is a top-bottom asymmetry in the refractive index of the cladding material. This results in the orthogonal polarization states to have slightly different effective refractive indices ($n_{eff}$). This results in dispersion offsets between the two polarizations and diminishes the on-chip throughput and spectral resolution of AWGs / Bragg-gratings / ring resonators, etc. Building a polarization-independent device requires careful planning and control of the fabrication processes \cite{smit1996phasar}. 

\noindent \textbf{4. Nanofabrication:} To achieve high coupling efficiency, high throughput and polarization independence requires careful planning of the chip material, fabrication recipe, and stable quality control. Similar process control is necessary for fiber-based devices. This highlights the need for dedicated nanofabrication facilities and expertise that should be available for astrophotonics research groups for rapid device development. 

\noindent \textbf{5. Detector technology:} It is possible to directly bond a 1D/2D detector array to the output facet of a photonic spectrograph, providing a triple benefit: 1. eliminating the chip-to-free-space interface, thus improving the overall throughput by as much as 30\%, 2. allowing stacking of photonic spectrographs to make a multi-object spectrograph (or IFU) with a single detector\cite{harris2012applications} , and 3. making the setup extremely stable, compact, and packageable. The upcoming detectors such as superconducting, photon-counting, energy-resolving microwave kinetic inductance detectors (MKIDS) may be packaged with the photonic-chip, and the whole assembly can be cooled cryogenically. A small pixel pitch of a few $\mu$m's is a critical requirement for direct bonding to an astrophotonic spectrograph chip to fully utilize their spectral resolution.   

\noindent \textbf{6. Mid-IR band:} As described earlier, the star-planet contrast is more favorable in the mid-IR (4-10 $\mu$m). In addition, optical interferometry is something that is possible with an astrophotonic instrument. So far, the efforts have been concentrated around the near-IR (J and H bands) due to the legacy from telecommunication technology. ZBLAN\footnote{Fluoride glass with a composition ZrF$_4$-BaF$_2$-LaF$_3$-AlF$_3$-NaF} fibers are a promising new development for mid-IR fiber feeds (GRAVITY\cite{perraut2018single} instrument is using ZBLAN fibers to ensure the best throughput in the K-band). Development of broadband mid-IR astrophotonic beam combiners is already well underway using chalcogenide glass platform \cite{arriola2017mid, tepper2017ultrafast}. Similar efforts need to be undertaken in visible wavebands as well. 

We suggest a two-fold strategy to address these issues: Technological (\S\ \ref{subsect:Technological_strategy}) and Organizational (\S\ \ref{sec:Organizational}).

 \begin{figure} [ht]
  \begin{center}
   \begin{tabular}{c} 
   \includegraphics[height=10cm]{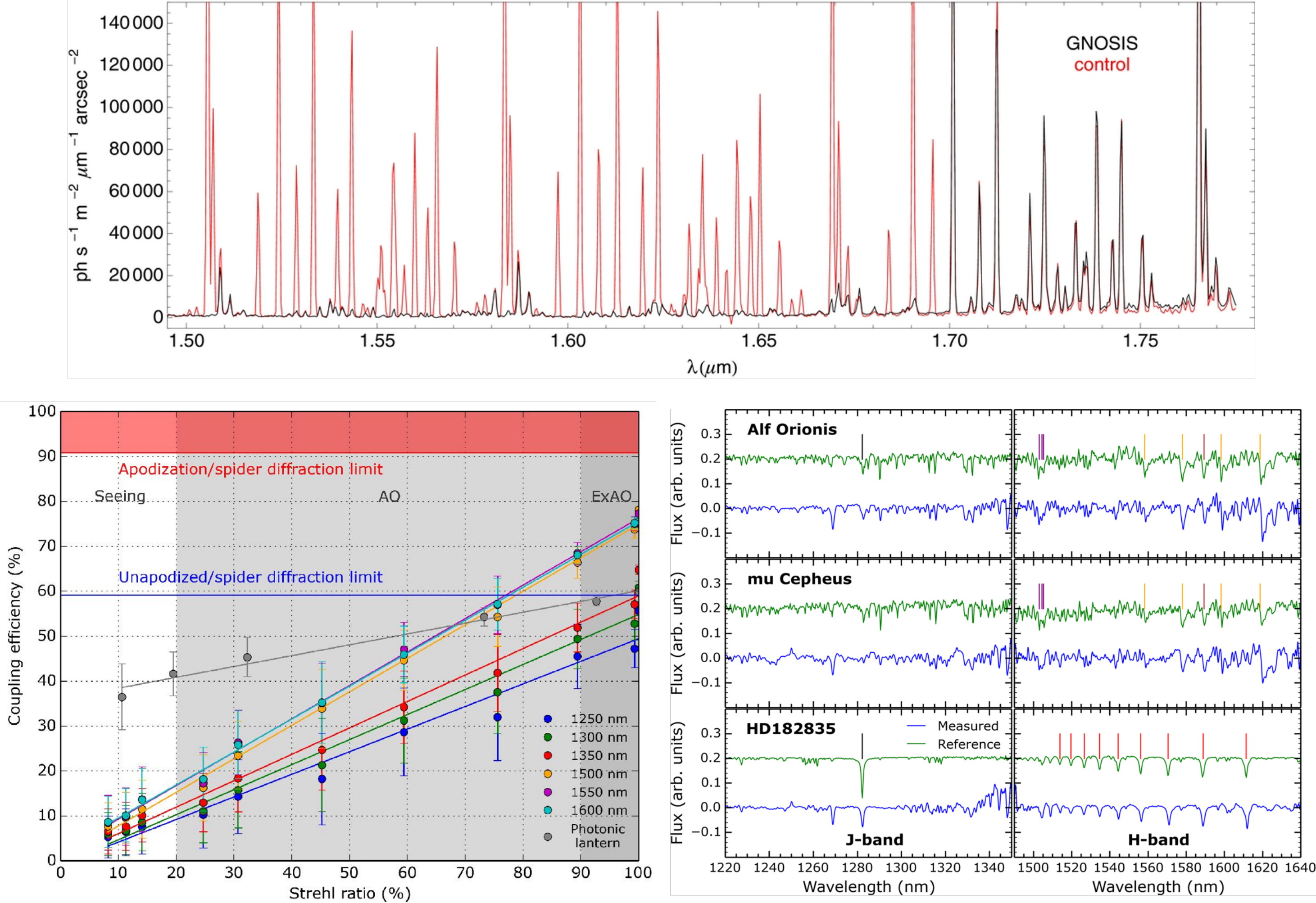}
   \end{tabular}
   \end{center}
   \caption[] 
   { \label{fig:On_sky_photonics} 
On-sky demonstrations of key astrophotonic technologies. a. Atmospheric OH suppression by fiber Bragg gratings \cite{trinh2013gnosis}, b. Coupling of extreme AO to a single-mode fiber (SMF) \cite{jovanovic2017efficient}, c. integrated photonic spectrograph using an SMF-fed AWG \cite{jovanovic2017demonstration}}
 \end{figure}

\subsection{Technological Strategy} \label{subsect:Technological_strategy}

To make astrophotonic instrumentation science-ready, we need to achieve a throughput comparable to conventional optics. For a preliminary calculation, we can compare the Echelle spectrograph in the X-shooter NIR arm \cite{vernet2011x} with the SCExAO single-mode-fiber-fed AWG spectrograph \cite{jovanovic2017demonstration} having similar resolving power of R $\sim$ 5000. Their total throughputs peak at 30\% for X-shooter and 5\% for SCExAO with 65\% Strehl ratio (13\% throughput with AR coatings and better alignment) at 1550 nm. This sets a goal of doubling the throughput of the photonic spectrograph system. Here, we also need to take into account other benefits of photonic spectrographs (eg: stability, compactness,  atmospheric OH-emission suppression, etc). 

\noindent With this picture in mind, the following technical improvements need to be undertaken to bring the throughput of photonic instruments on par with the conventional analogs. 

\noindent \textbf{1. Coupling efficiency:} All of the following options need to be exercised to optimize the coupling efficiency in/out of a photonic system:

\setlength{\leftskip}{0.5cm}
\noindent \textbf{A.} \textbf{Fiber - waveguide - free space:} It is important to match the numerical apertures (NA) at the interfaces to maximize the coupling efficiency. Fiber-waveguide coupling efficiency is the easiest target. About 95\% coupling efficiency has recently been demonstrated between a high numerical aperture single-mode fiber and an on-chip SiN waveguide \cite{zhu2016ultrabroadband, gatkine2016development} by employing an on-chip adiabatic taper which slowly changes the mode profile of the waveguide to match that of the fiber. A similar taper needs to be designed for waveguide to free space interface.  

\noindent \textbf{B.} For the \textbf{coupling between adaptive optics and single-mode fibers}, high-efficiency photonic lanterns can be used to couple the few-moded light into a set of single-mode fibers (throughput $\sim$ 85\%, \cite{trinh2013gnosis}). Further effort in this direction is desirable. As the extreme-AO evolves to achieve $>$ 90\% Strehl ratios, the number of modes required will be smaller, thus increasing the coupling throughput to a single or few-moded fiber (See Fig.  \ref{fig:On_sky_photonics}). We advocate for a continued effort in achieving and improving extreme-AO capabilities in large observatories. 

\noindent \textbf{C.} There is a need to develop \textbf{all-fiber NA converters} that conserve {\'e}tendue to achieve high coupling efficiency to both the extreme-AO as well as photonic chip\cite{leon2017divide}. 

\noindent \textbf{D.} By using \textbf{integrated (on-chip) photonic lanterns}, the fiber-chip interface is eliminated altogether. Work on this is underway at AAO and Univ. of Sydney, Australia \cite{Cvetojevic2017}. 

\setlength{\leftskip}{0pt}

\noindent \textbf{2. On-chip throughput:} \textbf{A.} Over the past few years, silicon nitride (SiN) platform has emerged as a low-loss photonic platform in visible, near- and mid-IR wavelengths \cite{munoz2017silicon}. It is possible to achieve $\sim$90\% on-chip throughput in photonic devices on SiN platform provided a tight process control is exercised in fabrication \cite{blumenthal2018silicon}. Such control with nanoscale precision can be achieved using stable electron-beam / extreme-UV lithography. It is crucial to have access to these and other state-of-the-art SiN fabrication facilities $+$ expertise to exploit this platform in astrophotonics. \textbf{B.} Another technique called ultra-fast laser inscription (ULI) is used for writing $\mu$m-scale structures in glass substrates \cite{thomson2009ultrafast}. Thanks to the one-step fabrication process, this is a quick, scalable method with easier process control or repeatability. This technique has been demonstrated for building optical, NIR, and mid-IR devices \cite{butcher2018ultrafast}. It is also important to devote resources to this technique to improve the throughput and enable a complete photonic device on a single platform.  

\noindent \textbf{3. Polarization dependence:} One way to resolve the polarization sensitivity problem is to split the linear polarizations \textit{a priori} and use two photonic chips/devices \cite{chen2011compact}. In addition to solving the problem, this will also provide a polarization measure for astrophysical observations such as GRB afterglows. However, splitting may not always be desirable due to signal-to-noise concerns. A polarization-independent astrophotonic chip can be achieved using a tightly controlled fabrication process to ensure uniform density of cladding around a square waveguide or by precisely controlled waveguide geometry to eliminate the offsets due to polarization sensitivity \cite{milovsevic2008design, dell2018novel}. ULI fabrication is another avenue for making polarization-independent devices. 

\noindent \textbf{4. Mid-IR waveband:} Further development of the  4$-$10 $\mu$m waveband astrophotonic devices will require focus on fabrication of customized mid-IR fibers (eg: ZBLAN fiber platform) and improvements in mid-IR fabrication processes (eg: on chalcogenide glass platform) to achieve stability over larger footprints ($\sim$ few tens of cm$^2$) since the chip footprint scales with wavelength. 

\noindent \textbf{5. Balloon-based / Cubesat tests:} The compact astrophotonic devices are ideally suited for the weight-conscious balloon and cubesat missions. Various innovative ideas such as space-based optical/IR interferometry \cite{lacour2014cubesats} can be tested using balloon or cubesat missions at relatively low cost, which would become pathfinders for future space telescopes.  

\noindent On-sky prototyping and validation of new and exciting astrophotonic approaches is essential prior to instrument deployment on ELTs. Current large U.S. telescopes (Keck, Gemini, Magellan) and continued development of their extreme-AO platforms are therefore of paramount importance for realizing the potential of astrophotonic instruments on the ELTs.

\section{Organizational Strategy} \label{sec:Organizational}

It is noteworthy that networks of institutions in Australia and Europe have taken major steps forward in the development of astrophotonics by enabling platforms fostering communication and collaboration with the industry and other disciplines, thereby increasing the accessibility of cutting edge technologies to astronomy and serving the diverse community back with new innovations. As a result, the key milestones in the growth of astrophotonics have come from Europe and Australia. With the availability of a large pool of photonics expertise in the American industry, the US astronomy community needs a renewed focus on astrophotonics to utilize this technological potential and become a key contributor in the coming decade. Our central idea is to create a distributed, multi-disciplinary \textbf{Institute of Astrophotonics} to streamline the development of science-ready astrophotonic devices. This idea is based on the successful models of innoFSPEC in Germany (\url{https://innofspec.de/en}) and CUDOS in Australia (\url{http://www.cudos.org.au/}). 

\noindent \textbf{innoFSPEC} is a research and innovation center pursuing multidisciplinary research in the field of optical fiber spectroscopy and sensing. It is a joint venture of the Leibniz Institute for Astrophysics Potsdam (AIP) and the Physical Chemistry group of the University of Potsdam. With access, international collaboration, and guaranteed observing times on various ground-based telescopes, innoFSPEC is well-placed for on-sky tests of astrophotonic instruments.

\noindent \textbf{CUDOS} is an Australian Research Council (ARC) Center of Excellence. It has brought together a powerful team of Australian and International researchers in optical science and photonics technology to lead significant advancement in capabilities and knowledge in this crucial field. It works closely with the Australian National Fabrication Facility (ANFF: \url{http://www.anff.org.au/}) which provides access to state-of-the-art fabrication technology and expertise. 

\noindent The requirements of astrophotonic instruments are non-conventional and therefore often requires pushing the technologies to their limits to obtain the best performance. This is true in the case of custom fabrication processes requiring nanoscale precision over a few cm$^2$, characterization requiring alignment within tens of nanometers, as well as packaging. While the cost of reproducing a device is low, the development of these devices require state-of-the-art facilities. 

\noindent The suggested \textbf{Institute of Astrophotonics} will be a network of institutes, universities / departments, and national fabrication facilities in the US (eg: NIST, Argonne National Lab) which will work closely together on developing the next generation of innovative astrophotonic instruments. This network will provide dedicated access to various facilities and expertise in fabrication, characterization, and packaging which will greatly shorten the development time and accelerate the delivery of science-ready photonic and astrophotonic instruments. 

\noindent Currently, the University of Maryland is actively developing on-chip astrophotonic instruments from design to fabrication to characterization. Some of the other US institutions involved in astrophotonic instrumentation include NASA/Goddard, NASA/JPL, Univ of Michigan, Caltech, and more. There is exciting work going on in photonics, especially at Univ. of California Santa Barbara on the SiN platform. NIST and Argonne National Lab (ANL) have advanced fabrication and characterization facilities. ANL has an ongoing collaboration with Macquarie University in Australia on astrophotonic applications of ring resonators. 

\noindent \textit{\textbf{However, there is an urgent need for a platform to bring all the stakeholders together and enable sharing of expertise and resources to meet the astrophotonics requirements for the next decade. Not only astronomy, but several other disciplines such as medical diagnostics, environmental analysis, sensing, and telecommunication industry will also benefit from such an institute.}}

\bibliography{report} 
\bibliographystyle{spiebib} 

\end{document}